**Tuning the Optoelectronics of Mixed-Semiconductors through the interplay of Quantum confinement and Stoichiometry Engineering**

*Kanha Ram Khator[1,3], Anupam Manna[2,3], Amlandeep Nayak[1], Pravat Nayek [2,3], Prasenjit Mal[2,3*], Satyaprasad P Senanayak[1,3*]*

[1] Nanoelectronics and Device Physics Lab, School of Physical Sciences, National Institute of Science Education and Research, (NISER), Bhubaneswar 752050, Odisha, India

[2] School of Chemical Sciences, National Institute of Science Education and Research, (NISER), Bhubaneswar 752050, Odisha, India

[3] Homi Bhabha National Institute, Training School Complex, Anushakti Nagar, Mumbai 400094, India

*Authors to whom correspondence should be addressed: satyaprasad@niser.ac.in, pmal@niser.ac.in

## Abstract

All-inorganic cesium lead bromide ($CsPbBr_3$) nanocrystals (NCs) have established themselves as an emerging semiconductor for next-generation optoelectronic technologies due to their unique combination of properties, such as near unity photoluminescence quantum yields, narrow color pure emission, and exceptional defect tolerance. Although size-dependent optical signatures of these NCs are well reported, the complexity of mixed ionic-electronic transport remains largely unexplored. In this study, we provide a comprehensive analysis of size-dependent charge transport by decoupling ionic and electronic transport dynamics through carefully designed transient current and space charge limited current measurements. By employing NCs of different sizes ranging from 5.6 nm to 11.3 nm in thin films, we provide a comprehensive understanding of quantum confinement effects and related synthetic chemistry. Contrary to popular beliefs of quantum confinement and band gap broadening, our results demonstrate that the smallest NCs exhibit the most efficient transport characteristics, evidenced by the lowest activation energy ($E_a^{hole}$= 78 meV) for hole transport and the highest barrier for vacancy-mediated ion migration ($E_a^{ion}$= 370 meV). This work paves a way forward for perovskite-based efficient quantum devices, by demonstrating that moving into a strong quantum confinement regime, a superior charge transport can be facilitated, when supported by carefully tailored stoichiometry.

## 1. Introduction

Integrating NCs into thin films is a promising approach for creating new solution-processable semiconductor-based devices.[21, 54, 57] Perovskite NCs, specifically $CsPbBr_3$ NCs, hold special attention, due to their exceptional optoelectronic properties, which can be precisely modulated, thereby opening up a plethora of applications.[19, 39, 44] This advantage of $CsPbBr_3$ stems from the tunability of the NC's size[32, 39], shape[17, 27], composition[12], surface chemistry[20, 52], and spatial arrangement[41]. Among these tunable attributes, size and/or facets conventionally govern the quantum confinement effects and the associated electronic band structure, which eventually impacts the semiconducting properties of this class of materials.[11, 20, 34, 39] Conventionally, smaller NCs within the strong-confinement regime exhibit a larger band gap and a blue-shifted photoluminescence (PL) peak in comparison to larger nanocrystals.[5, 33, 39, 46] Specifically, in the case of $CsPbBr_3$ nanocrystals, the confinement effect modulates the density of states in the valence band edge, resulting in a significant size-dependent Stokes shift.[8] Furthermore, $CsPbBr_3$ NCs have also shown size-dependent bi-exciton lifetimes (35-200 ps) following a positive correlation with size.[4] Notably, these nanocrystals (NCs) exhibit a non-monotonic size dependence[1], reflecting a complex interplay of quantum confinement[8], the excitonic density of states[16, 47], and size-driven structural transitions[8, 9]. In halide perovskite NCs, variation in size also alters lattice symmetry, often inducing a transition from the cubic to the orthorhombic phase.[50] These observations collectively underscore the critical role of NC size in shaping the optical density of states and the associated energy landscape. Despite this, systematic studies directly correlating nanocrystal size (or edge length) with both electronic and ionic transport remain limited. Establishing such relationships is essential for developing a comprehensive understanding of charge and ion dynamics in perovskite semiconductors, particularly in the context of device miniaturization approaching the limits imposed by Moore's law.[26, 53]

Versatility in the synthesis of $CsPbBr_3$ nanocrystals has allowed precise control on the size, shape, defect density and optoelectronic properties of nanocrystals.[44] Conventional hot injection method for NC synthesis offers reasonably precise control over nucleation and growth kinetics, enabling the formation of largely monodisperse nanocrystals with higher degree of crystallinity.[39] However, this synthetic process does not inherently eliminate defect formation. Specifically, the method of rapid precursor injection at elevated temperatures leads to a spike in nucleation followed by fast growth, resulting in the formation of kinetically trapped defects—particularly halide vacancies and surface under-coordination.[10] In addition, it has also been observed that the higher reaction temperature can promote ligand desorption or dynamic ligand binding, resulting in imperfect surface passivation and generation of defects or non-radiative recombination centers.[13, 22, 45] Moreover, the fast nucleation growth in the hot-injection methods also limits fine control over precursor stoichiometry during growth, thereby disturbing the overall stoichiometry and contributing to defect formation. Other techniques, such as ligand-assisted re-precipitation (LARP)[44] or mechanochemical[40] and solvent-free approaches[18] although simplify synthesis, they typically result in poly-dispersed particles with a lower degree of crystallinity and higher trap densities. These variations in synthetic accessibility directly influence halide vacancy concentration, surface defect density, and size dispersion, which in turn govern charge carrier dynamics, ion migration, and film formation. Consequently, synthesis-dependent structural and compositional heterogeneities must be carefully optimized to balance scalability with the electronic and ionic transport properties required for high-performance perovskite devices.

In this study, we utilize precursor engineered NCs of $CsPbBr_3$ to probe the underlying electronic and ionic transport in different field regimes. We observe that modification of the precursor for synthesis systematically modulates the NC size/edge length to vary by two times, ranging from 11 nm to ~ 5 nm, thereby affecting the energetic band-structure. Comprehensive charge transport measurements provide conclusive evidence that the smallest NCs, despite a higher band gap and stronger confinement, exhibit superior electronic properties with a hole mobility ($\mu_h$) ~ $1 \times 10^{-3}$ $cm^2/Vs$, orders of magnitude higher than the NCs with larger sizes. Simultaneously, we also observe a decrease in the ionic defect migration in the smallest NCs. These trends are primarily attributed to the impact of precursor-chemistry in slowing down the nucleation process, thereby resulting in monodispersed NCs with an extremely low degree of defect density.

## 2. Experimental Methods:

*Synthesis of $CsPbBr_3$ NCs:* $CsPbBr_3$ NCs of varying sizes were synthesized using a modified hot injection method.[37] *t*-$CsPbBr_3$ and *d*-$CsPbBr_3$ were synthesized using tribromoisocyanuric acid (TBIA) and dibromoisocyanuric acid (DBIA), respectively, as the bromine source, along with PbO as the lead precursor and cesium oleate as the cesium source. *p*-$CsPbBr_3$ was synthesized using $PbBr_2$ as the precursor for both lead and bromine source for a comparative study. In the synthesis of all three NCs, oleic acid (OA) and oleylamine (OLA) were employed as capping ligands and octadecene (ODE) as the non-coordinating solvent. The reactions were quenched in an ice bath, followed by purification using methyl acetate (MeOAc) and re-dispersion in hexane. Elaborated synthesis procedures for all the three NCs have been provided in the supplementary note 1.

*Fabrication of 2-terminal charge transport devices:* 2-terminal devices (ITO/PEDOT:PSS/$CsPbBr_3$/Au) were fabricated for all the three NCs on precleaned (ultrasonication in soap solution/DI water/Acetone/IPA followed by UV Ozone treatment) patterned ITO substrates. PEDOT:PSS (from Ossila) was spin-coated at 3000 rpm for 30 seconds, followed by annealing at 130 °C for 15 minutes in ambient conditions. The substrates were then transferred to a $N_2$ filled glove box ($O_2$ < 0.1 ppm, $H_2O$ < 0.1 ppm) for further processing. $CsPbBr_3$ films were obtained by spin coating (4000 rpm, 60 s) the colloidal suspension of NCs in hexane (10 mg / 400 µL), followed by annealing at 90 °C for 30 minutes. Top Au electrodes of 25 nm thickness were deposited using a Moorfield Minilab 026 Thermal Evaporator at a deposition rate of 0.1 Å/s with a chamber pressure of ~$10^{-6}$ mbar.

*Time of Flight mobility estimation of the NCs:* The transit time corresponding to the time of flight measurement was estimated from the 10% – 90% rise time ($\tau_{rise}$) of the transient current signals, and the time of flight mobility ($\mu_{TOF}$) was calculated according to the following relation: $\mu_{TOF} = \frac{L^2}{V\ \tau_{rise}}$; Where L is the thickness of the thin film, and V is the applied bias. The energetic barrier for this transport process was estimated from temperature-dependent mobility using an Arrhenius dependency of activation energy.

*Transient current measurements:* Transient current measurements were performed for the estimation of the ionic activation energy ($E_a^{ion}$) of NCs; Square pulses of 100 Hz (**Figure 1a**) were applied to the device using Teledyne T3AFG120 function generator, and the response of the device was recorded using Teledyne WavePro 404HD Oscilloscope. The electrical circuit utilized in the measurement is provided in inset of **Figure 1a**. Temperature-dependent transient current measurements were performed using a closed-cycle liquid helium cryostat integrated with Lakeshore Probe Station (CRX 6.5K).

*Space Charge Limited Current (SCLC) Measurements:* All J-V characteristics were performed using a Lakeshore probe station and Keysight B1500A Semiconductor Device Analyzer. Temperature-dependent J-V measurements were measured in a range of 100 K - 300 K using a closed-cycle liquid helium cryostat integrated with Lakeshore Probe Station (CRX 6.5K).

## 3. Result and Discussion:

Nanocrystals (NCs) of $CsPbBr_3$ were synthesized using the standard hot-injection method and are denoted as *p*-$CsPbBr_3$. For comparative analysis, two additional variants—*t*-$CsPbBr_3$ and *d*-$CsPbBr_3$ were prepared utilizing a set of organo-bromide precursors, namely tribromoisocyanuric acid (TBIA) [14] and dibromoisocyanuric acid (DBIA) [31], respectively. Although TBIA and DBIA possess nearly identical chemical structures, they differ in their bromide content per molecule, with TBIA containing approximately 33% more bromide than DBIA. Interestingly, with this variation in precursor composition, a systematic change in nanocrystal size is observed. The average particle sizes were determined to be 11.3 ± 2.1 nm for *p*-$CsPbBr_3$, 8.3 ± 0.6 nm for *d*-$CsPbBr_3$, and 5.6 ± 0.4 nm for *t*-$CsPbBr_3$. Detailed synthesis procedure and relevant references are provided in the methods section, **Supplementary Note 1,** and are illustrated schematically in **Figure S1**. Notably, this precursor-engineering strategy not only enables precise control over nanocrystal size but is also expected to modulate the density of halide (bromide) vacancies within the NCs. To investigate the impact of these variations on charge transport, vertical device structures were fabricated with the architecture: ITO/PEDOT:PSS/$CsPbBr_3$/Au, enabling the study of both electronic and ionic transport properties. The devices were designed to preferably facilitate hole transport, as Br-based perovskites have been reported to exhibit dominant p-type behavior.[35, 43, 51, 57] Detailed description of the device fabrication is provided in the Methods section. Transient measurements were first performed on these vertical devices using a square voltage pulse (5 V, 100 Hz). The resulting transient current was monitored via the voltage drop across a load-matched 1 MΩ resistor connected in series with the device (**Figures 1a** and **S2**). Since these perovskite semiconductors exhibit mixed ionic–electronic conduction, a relatively low input frequency (~100 Hz) was used to ensure that the measured transient response accurately captured the combined charge-transport dynamics. At such frequencies, the slower ionic species have sufficient time to respond to the applied signal, resulting in a transient response that represents a convolution of both electronic and ionic transport processes within the nanocrystals. Upon application of the square pulse, the transient current exhibits an exponential rising temporal feature before reaching a steady-state value. To ensure that this temporal response has no contribution from parasitic capacitance associated with probes or connectors, we utilized frequency-matched probes ~ up to 1 GHz for these measurements. We first estimated the rise time ($\tau_{\mathrm{rise}}$) defined as the time taken for the channel current to increase from 10% to 90% of its maximum value associated with the transient response. It is observed that $\tau_{\mathrm{rise}}$ decreases from ~46 μs for *p*-$CsPbBr_3$ to ~35 μs for *t*- $CsPbBr_3$. Based on these rise times, the time-of-flight mobility ($\mu_{TOF}$) of holes was estimated to be in the range of $10^{-4}$ to $10^{-7}$ $cm^2\ V^{-1}\ s^{-1}$. This mobility predominantly originates from nearest-neighbor hopping of holes through the HOMO levels of the nanocrystals (**Figure 1d**).[30] However, this mobility would also have an impact of the ionic defects which impede hole transport.[42] Notably, $\mu_{TOF}$ exhibits a general trend indicating an increase in $\mu_{TOF}$ with a decrease in nanocrystal size (**Figure S3**). However, it should be stated that the saturation current density ($J_{tr}^{sat}$) values were obtained to be 43.3 $mA/cm^2$,14.7 $mA/cm^2$, and 13.8 $mA/cm^2$ for *p*-, *d*-, and *t*-$CsPbBr_3$ based devices at 300 K respectively. The variation in $J_{tr}^{sat}$ appears to be inconsistent with the trends in $\mu_{TOF}$. However, this apparent discrepancy originates from the substantial variation in film thickness, L ~ (100 ± 16) nm, (423 ± 22) nm, (490 ± 25) nm for *p*-, *d*-, and *t*-$CsPbBr_3$, respectively. Under an applied bias $V$, the internal electric field scales as $E = V/L$, implying that thicker films experience significantly lower effective fields. In nanocrystal solids where charge transport occurs via thermally activated nearest-neighbour hopping, the carrier mobility exhibits a strong field dependence, often described by Poole–Frenkel-type behavior $\mu\,(E) \propto \exp(\gamma\sqrt{E})$.[7, 28, 38] Additionally, in the high-bias regime, transport approaches space-charge-limited conduction (SCLC), where the current density follows $J \propto \mu(E)\,V^2/L^3$ (described in details in later section). Consequently, both the explicit $L^{-3}$ scaling and the implicit dependence of $\mu(E)$ amplify the sensitivity of $J_{tr}^{sat}$ to film thickness.[2] Therefore, a direct comparison of $J_{tr}^{sat}$ across devices with different thicknesses is not physically meaningful. Upon normalizing for thickness (i.e., accounting for the effective field and geometric scaling), the corrected $J_{tr}^{sat}$ was obtained to be 8.9 mA/cm$^2$, 12.7 mA/cm$^2$, and 13.8 mA/cm$^2$ for *p*-, *d*-, and *t*-$CsPbBr_3$ respectively, which is consistent with the trends in $\mu_{TOF}$.

To further elucidate the charge-transport mechanism, the transient current response was quantitatively fitted over the exponential relaxation regime up to the steady-state equilibrium channel current. The transient response can be well described using an exponential form $I(t) = I_\infty\left(1 - e^{\frac{t}{\tau_c}}\right)$, (where $I_\infty$ is the equilibrium steady state current), consistent with first-order relaxation kinetics.[25] In this asymptotic regime, the characteristic time constant ($\tau_c$), combines the coupled effect/impact of slower ionic relaxation over fast electronic transport, thereby providing a comprehensive description of the mixed-transport behavior. In principle, hole transport is expected to be impeded by ionic screening arising from the accumulation of mobile ions, resulting in a current decay. However, the accumulation of ions ($Br^-$) at the metal-semiconductor interface also progressively reduces the effective electric field, which manifests as a delayed transient response (**Figure 1d**). Under an applied bias, holes are injected across the metal-perovskite interface over an energy barrier of $E_{bh2}$. Concurrently, anionic species—most plausibly $Br^-$ ions migrating via bromide vacancies ($V_{Br}$), drift and accumulate near the semiconductor–metal (anode) interface[15]. This agrees well with the calculation showing that $V_{Br}$ exhibits the lowest activation energy for vacancy-assisted diffusion among the other dominant defects/vacancies, i.e., $Cs^+$ and $Pb^{2+}$ cations. [24, 25, 57] The field-driven redistribution of $V_{Br}$ effectively modulates the local electrostatic potential landscape, thereby inducing a band-bending. As a result, the interfacial injection barrier decreases (to $E_{bh3}$), thereby facilitating hole transport despite the concurrent screening effects. A schematic depicting the band-bending at the anode-perovskite interface due to defect/vacancy accumulation, as well as the applied bias, is shown in **Figure 1e-g**. Notably, the transient current exhibits a single dominant characteristic time scale, indicating that $Br^-$ migration and the associated modulation of the hole injection barrier govern the transient asymptotic response[25]. The contribution of heavier ionic species is therefore negligible under the present experimental conditions. At room temperature, $\tau_C$ was estimated to be ~ 19.7 μs and ~ 17.8 μs for *p*-$CsPbBr_3$ and *t*-$CsPbBr_3$, respectively. These values are slightly lower than $\tau_{rise}$ since the transient current in this regime is governed by the hole transport under a lower injection barrier induced due to ion accumulation at the metal-semiconductor interface. Moreover, $\tau_C$ is a quantitative fit for the signal transition from 100 % to ~ 63 %, while $\tau_{rise}$ was estimated from the 10–90% signal transition.

The activation energy associated with the ion migration across different NCs was estimated from the temperature-dependent transient current measurements. The saturation current density, $J_{tr}^{sat}$ decreased markedly with temperature, from ~43 mA/cm$^2$ at 300 K to ~3 mA/cm$^2$ at 100 K for *p*-$CsPbBr_3$, as shown in **Figure 2a**, indicative of a thermally activated charge transport behavior. Similar decreasing trend of $J_{tr}^{sat}$ was observed for *d*-$CsPbBr_3$ (14.5 to 1.5 mA/cm$^2$) and *t*-$CsPbBr_3$ (13.8 to 12.7 mA/cm$^2$) (**Figure 2b,c**), consistent with reduced ionic as well as hole mobility at lower temperatures. Unlike 3D perovskites, where lower temperatures minimize ionic screening to realize a negative coefficient of charge transport, here we observe an activated transport through the complete temperature range of measurement[42]. This indicates that the extent of ionic species is rather low in these class of materials. Nevertheless, the decrease in $J_{tr}^{sat}$ with temperature is also accompanied by the slowing down of the transient dynamics, as reflected in the magnitude of $\tau_C$ (**Figure 2d-f**). For *p*-$CsPbBr_3$ based devices, $\tau_C$ increases from 19.7 μs at 300 K to 125 μs at 100 K with analogous trends for *d*-$CsPbBr_3$ and *t*-$CsPbBr_3$. The fitting parameters and the associated time constants are summarized in **Appendix 1** in SI. This slowing down of the transient dynamics is essentially a signature of reduced thermal activation of vacancy-mediated ion migration, leading to delayed establishment of equilibrium saturation transient current. Notably, these time constants are estimated over varied temperature ranges for different NCs. This is because, below a threshold temperature ($T_{th}$), the exponential rise in the transient current disappears, and the response closely follows the applied square pulse, however with a reduced magnitude of $J_{tr}^{sat}$, indicating effective freezing of mobile ions. The extracted values are ~ 260 K for *t*-$CsPbBr_3$, ~ 160 K for *d*-$CsPbBr_3$, and ~ 100 K for *p*-$CsPbBr_3$. Higher $T_{th}$ value for *t*-$CsPbBr_3$ based devices suggests a lower density of ionic defects, leading to earlier suppression of ionic motion. Assuming thermally activated hopping, the characteristic time constant follows an Arrhenius relation $\frac{1}{\tau_c} = K = K_0 \exp\left(\frac{-E_a^{ion}}{k_B T}\right)$ where $E_a^{ion}$ is the activation energy for ion migration and K is a rate constant associated with these processes.[25, 36]. We found that *t*-$CsPbBr_3$ has the highest ionic barrier with $E_a^{ion}$ = (370 ± 44) meV, while *d*-$CsPbBr_3$ and *p*-$CsPbBr_3$ exhibit a lower $E_a^{ion}$ of (262 ± 47) meV and (174 ± 49) meV, respectively, indicating that vacancy-assisted $Br^-$ migration is more prominent in *p*-$CsPbBr_3$ (**Figure 2g-i, S4**). In summary, the transient measurements clearly elucidate the influence of nanocrystal size and precursor engineering on both electronic charge transport and ionic dynamics in these systems. The observed dependence of $\mu_{TOF}$ on nanocrystal size indicates enhanced electronic coupling and reduced hopping barriers in smaller nanocrystals, while variations in precursor chemistry systematically modulate the density and mobility of halide vacancies. As a result, the measured transient device response reflects a convolution of faster electronic hopping and slower field-driven ionic drift/diffusion, establishing a coupled transport regime that dictates the overall device kinetics.

To probe electronic transport in the high-field regime (~ 200 V/μm), we utilized the space-charge limited (SCLC) formalism. Considering the inherent p-type semiconducting character of $CsPbBr_3$, and the device design, we majorly expect hole transport.[30, 57] J-V characteristics depicted in **Figure 3a-c** exhibit a clear transition from the Ohmic ($J \propto V$) to the SCLC regime ($J \propto V^2$), which persists down to 100 K, indicating low trap densities and efficient injection in these devices. The charge carrier mobility was estimated from the SCLC regime, using the Mott-Gurney law $J = \frac{9}{8} \epsilon_0 \epsilon_r \mu_h \frac{V^2}{L^3}$, where J is the current density, $\epsilon_0$ is the permittivity of vacuum, $\epsilon_r$ is the dielectric constant of the active material, V is the applied voltage, and L is the film thickness.[2, 6, 43] The estimated mobility values exhibit a strong nanocrystal size dependence, with values as high as $1 \times 10^{-3}$ $cm^2/Vs$ for *t*-$CsPbBr_3$, the smallest NC (5.6 ± 0.4 nm), which decreases to $1.6 \times 10^{-5}$ $cm^2/Vs$ for *d*-$CsPbBr_3$ and reaches a value as low as $6.4 \times 10^{-8}$ $cm^2/Vs$ for *p*-$CsPbBr_3$ (**Figure 3a-c**) with a nanocrystal size of 11.3 ± 2.1 nm. Temperature-dependent J-V measurements reveal a typical activated behavior with two different regimes (**Figure 3d** & **S5**). In high T regime (200 K – 300 K), the current density as well as mobility follows an Arrhenius type behavior with $\mu_h = \mu_0 \exp\left(\frac{-E_a^{hole}}{k_B T}\right)$ where $E_a^{hole}$ is the hole activation energy, $k_B$ is the Boltzmann constant, T is the temperature, and $\mu_0$ is the pre-exponential factor. The activation energy from these SCLC-based devices was estimated to be in the range of (78 ± 11) meV for *t*-$CsPbBr_3$, which increases by around two times (156 ± 24) meV for *p*-$CsPbBr_3$. However, at lower temperatures (100 K – 200 K) and higher electric fields (corresponding to the SCLC regime), the transport exhibits a typical Poole-Frenkel mechanism, reflecting barrier lowering in the presence of an electric field (**Figure 3d** & **S5**) and consequently a lower activation energy in the range of 20 – 40 meV.

## 4. Mechanism:

A summary of the activation energy for both hole transport and ionic defect is provided in **Figure 4**. Interestingly, we observe that the smallest nanocrystal (*t*-$CsPbBr_3$) exhibits the lowest activation energy for hole transport and the highest activation energy for ionic transport. Similarly, the largest NC (*p*-$CsPbBr_3$), exhibits the lowest ionic activation energy, indicating ease of vacancy-mediated ionic defect migration and the highest hole transport activation energy. This enhanced performance of *t*-$CsPbBr_3$ compared to its counterparts can possibly be rationalized from the interplay between precursor-limited nucleation kinetics and quantum confinement effects during thin-film assembly. Relatively lower solubility of TBIA precursor moderates the halide release to the reaction mixture, slowing down the early-stage reaction kinetics.[14] This process results in the growth of defect-suppressed nanocrystals.[3, 55] In contrast, conventional precursors with higher solubility result in faster growth of larger NCs via an uncontrolled Ostwald ripening process, often trapping defects like halide vacancies within the crystal lattice. The impact of the kinetics on the NC quality is reflected in the near-unity photoluminescence quantum yield (PLQY = 99%[14], 93%[31], 75%[49] for *t*-, *d*-, *p*-$CsPbBr_3$, respectively). In addition, the improved stoichiometry obtained from elemental composition extracted from EDS (Cs:Pb:Br = 1:1:4.5 and 1.08:1:3.26 for *t*- and *d*-$CsPbBr_3$), confirms the presence of a bromide-rich composition and vacancy passivation.[14, 31] Furthermore, we estimated the size dispersion(%) = $\frac{Standard\ Deviation}{Mean\ Size} \times 100$ of the NCs from earlier reported TEM images [14, 31, 49, 56] and observe that *t*-$CsPbBr_3$ has a size dispersion of ~ 7 %, in comparison *p*-$CsPbBr_3$ based NCs exhibit a size dispersion ~ 18 %.[23] A combination of the well-defined crystalline facets and uniform packing (owing to lower dispersion) in these strongly confined *t*-$CsPbBr_3$-based NCs supports overlap of individual wavefunctions, resulting in a quasi-continuous density of states, thereby decreasing the overall energetic barriers for hopping transport of charge carriers (holes).[29, 48] Conversely, poly-dispersed, defect-rich larger NCs (*p*-$CsPbBr_3$) exhibit significant carrier localization due to disrupted coherence and a higher degree of ionic defect densities. All in all, we demonstrate a counterintuitive phenomenon where smaller NCs exhibit enhanced electronic coupling and defect-suppressed lattice energetics owing to controlled precursor chemistry and chemical kinetics.

## 5. Conclusion:

In summary, this study provides insights into the fundamental charge transport properties of different-sized NCs. Our measurements thus indicate a strong coupling of the electronic transport with ionic defect vacancies in $CsPbBr_3$ NCs. Both of these factors are strongly affected by the precursor chemistry and the associated quantum confinement effects in the energetic landscape. We conclude that, despite stronger quantum confinement and band gap broadening, the smallest NC, *t*-$CsPbBr_3$, shows superior transport properties, assisted by healing of halide vacancies and possible wavefunction overlap at NC boundaries. Improved transport characteristics were further confirmed from the highest ionic activation barrier (370 meV) and lowest hole activation energy (78 meV) for *t*-$CsPbBr_3$. We demonstrate that strong quantum confinement, reduction in halide vacancies through halide-rich precursors, and controlled reaction dynamics can significantly enhance the potential of using perovskite NC in high-tech quantum devices.

## Acknowledgements:

SPS acknowledges funding support from DAE through projects: RIN:4001 and RNI:4011, Intensification of Research in High Priority Areas (IRHPA) from the ANRF (IPA/2021/000096), ANRF/ARG/2025/001573/CS, Royal Society for Newton International Fellowship. PM thanks the Department of Atomic Energy (DAE), Government of India, for financial support (grant number RIN 4002). KRK and AM acknowledge NISER for the fellowship and resources. PN thanks UGC (India) for the fellowship.

## AUTHOR DECLARATIONS

### Conflict of interest

The authors declare that they have no known competing financial interests or personal relationships that could have appeared to influence the work reported in this paper.

### Author contributions:

**KRK**: Device Fabrication, Data Curation, Writing - Initial Draft. **AM** & **PN**: Synthesis of NCs. **AN**: Data curation. **PM**: Supervision, Finalizing Manuscript. **SPS:** Conceptualization, Methodology, Supervision, Writing, Review & Editing, Finalizing Manuscript.

## DATA AVAILABILITY

The data that support the findings of this study are available within the article and its supplementary material.

**Figures and Captions**:

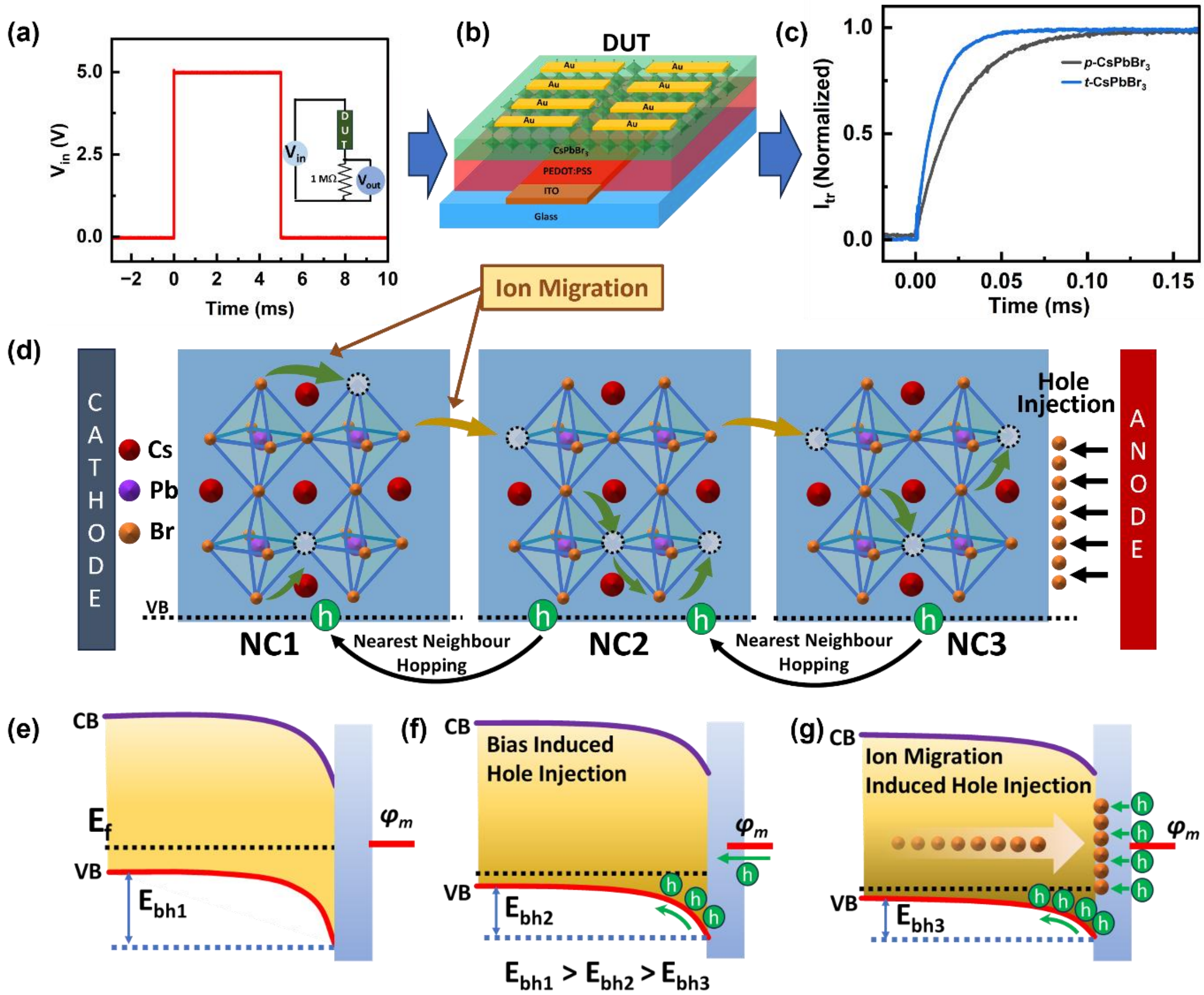

**Figure 1.** (a) Input square pulse for the transient current measurements, Inset shows the electrical circuit utilized for the measurement. (b) Schematic illustration of the device structure. (c) Representative Transient current density versus time graph corresponding to the input square pulse. (d) Representation of nearest neighbor hopping and ion migration through a NC array. (e) Band bending due to $E_f$ - anode work function alignment. $E_{bh1}$ represents the energy barrier for hole injection (f) upon biasing, holes are injected through the anode, and $E_{bh1}$ decreases to a value of $E_{bh2}$. (g) $Br^-$ accumulates at the anode interface due to ion migration, which enhances hole injection, further lowering the $E_{bh3}$.

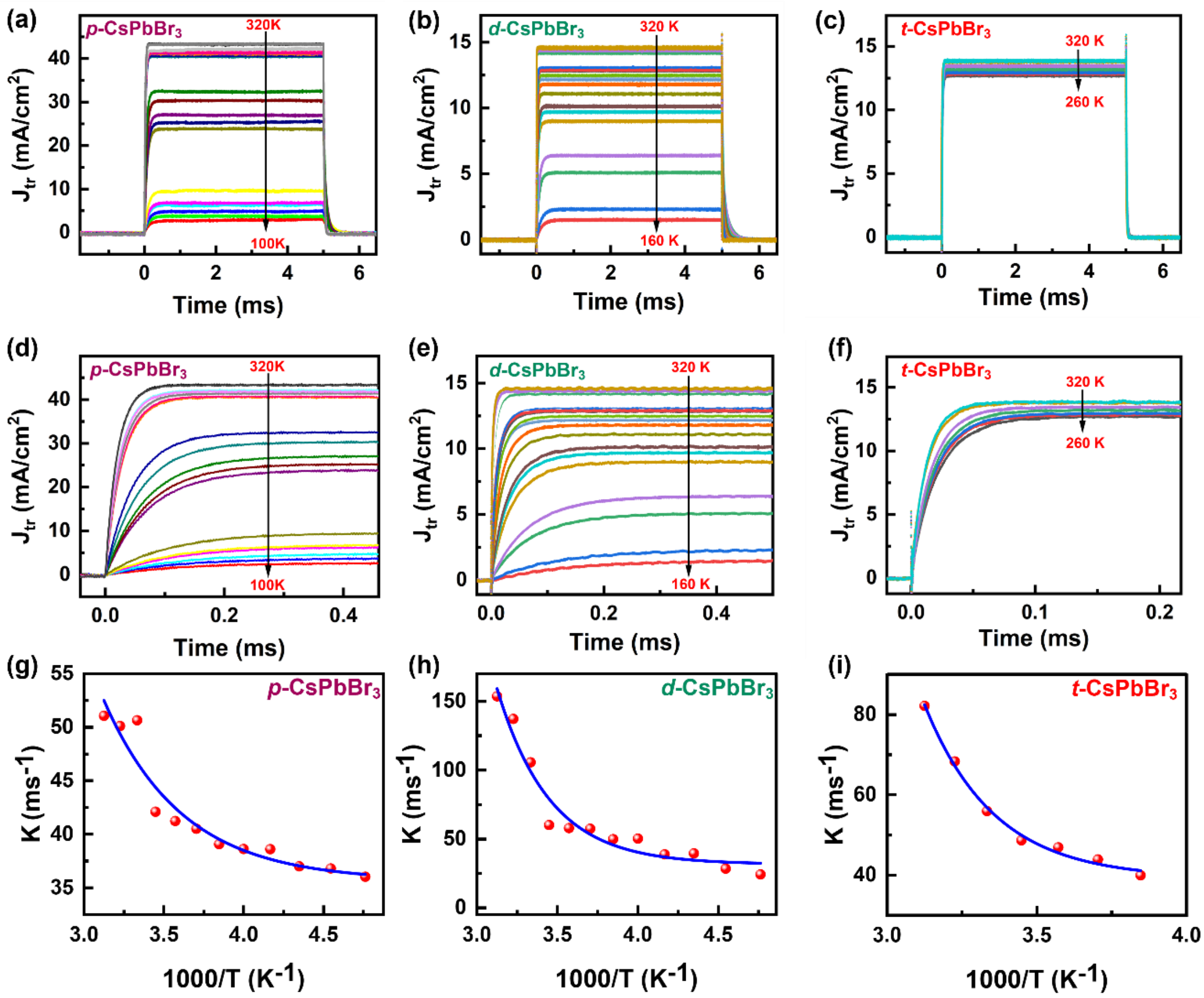

**Figure 2.** Temperature-dependent transient current measurement for (a) *p*-$CsPbBr_3$ (b) *d*-$CsPbBr_3$ (c) *t*-$CsPbBr_3$ NCs. (d, e, f) Zoomed in plots highlighting the exponential rise resulting from the intermixing of ion migration and hole transport. K versus 1000/T plot for (g) *p*-$CsPbBr_3$ (h) *d*-$CsPbBr_3$ (i) *t*-$CsPbBr_3$. Symbols are experimentally obtained time constant values, and the solid blue lines indicate the Arrhenius law fit.

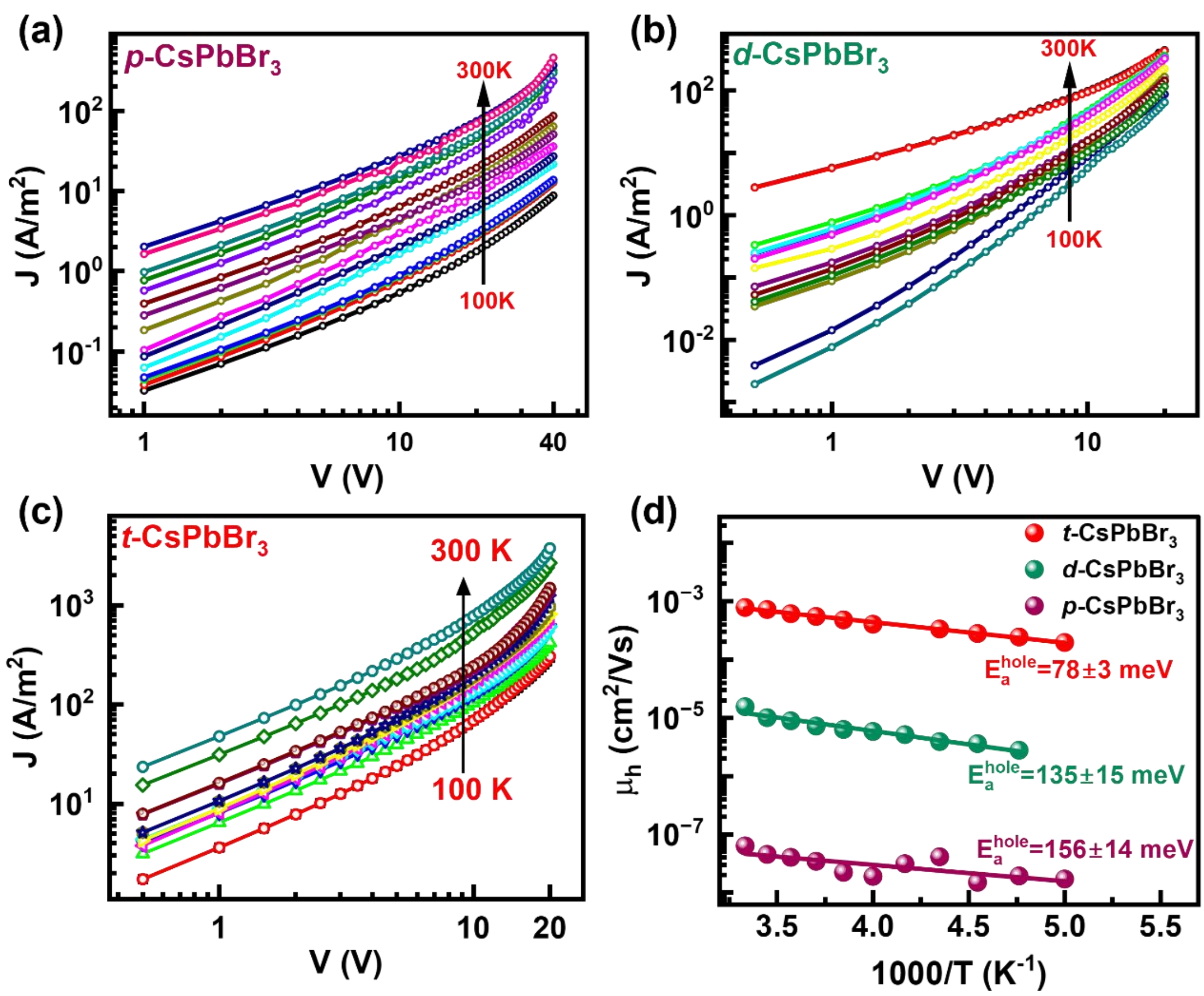

**Figure 3.** Temperature-dependent J-V characteristics for (a) *p*-$CsPbBr_3$ (b) *d*-$CsPbBr_3$ (c) *t*-$CsPbBr_3$ NCs. (d) Semi-log plots of $\mu_h$ Versus 1/T for all the NCs along with the respective activation energies.

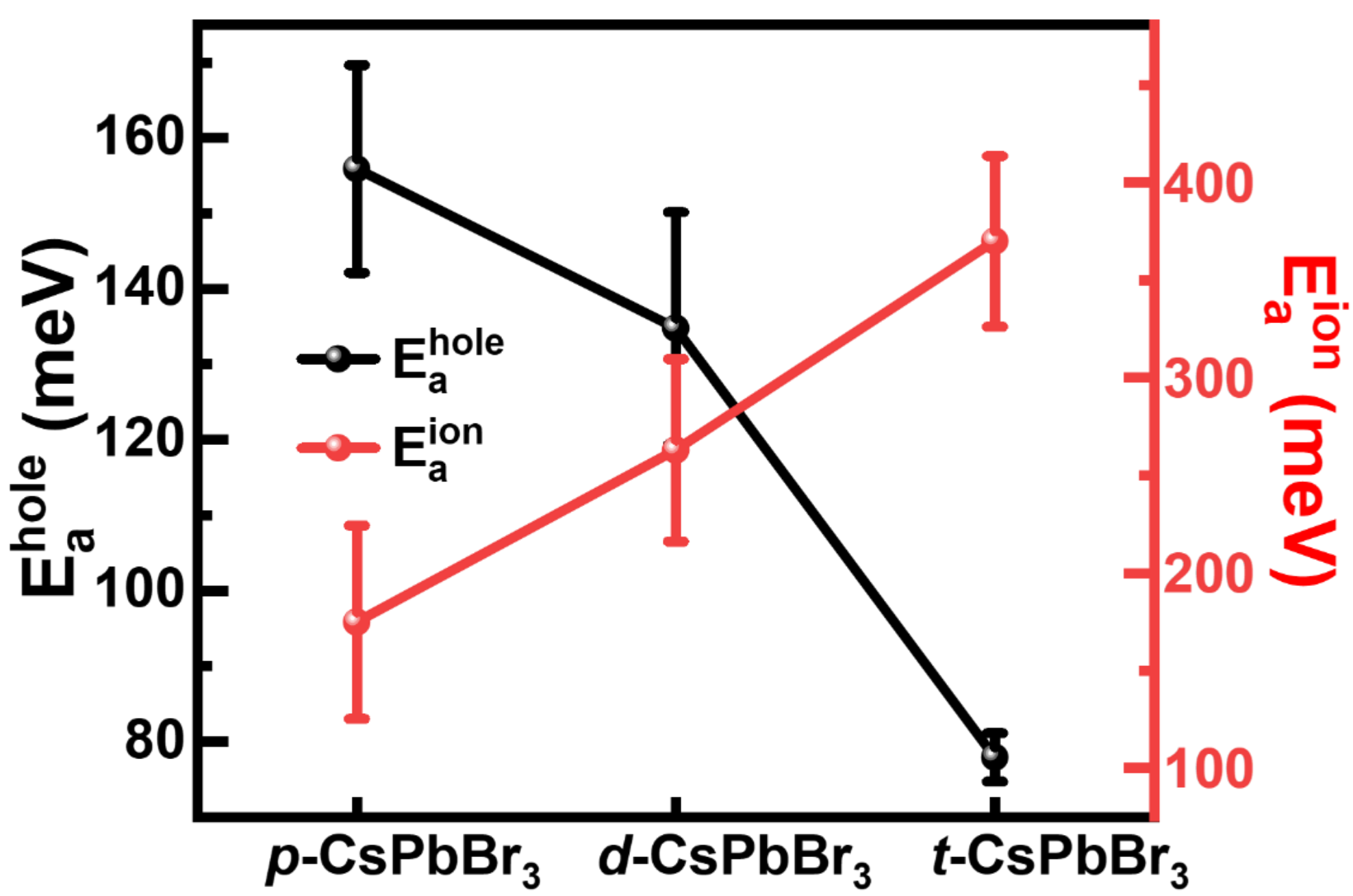

**Figure 4.** Comparison of $E_a^{hole}$ and $E_a^{ion}$ across all the studied NCs.